\pdfoutput=1

\documentclass[12pt,a4paper]{article}

\usepackage{ifthen} 
\usepackage{rotating}
\usepackage{dashrule}
\usepackage{array} 
\usepackage{dcolumn}
\usepackage{comment}
\usepackage{xcolor}

\usepackage{graphpap} 
\usepackage{rotating} 
\usepackage{tikz}
\usepackage{xcolor}
\usepackage{longtable} 

\usepackage{lscape}
\usepackage{multirow} 
\usepackage{mathrsfs}
\usepackage{mathtools} 
\usetikzlibrary{patterns}
\usepackage{nicefrac} 
\usepackage{transparent}
\usepackage{afterpage} 
\usepackage{comment} 
\usepackage[normalem]{ulem}
\usepackage{cancel} 

\newboolean{pdflatex}
\setboolean{pdflatex}{true} 

\newboolean{articletitles}
\setboolean{articletitles}{true} 

\newboolean{uprightparticles}
\setboolean{uprightparticles}{True} 

\def\paperauthors{T.~Ovsiannikova on behalf of LHCb collaboration} 
\def\paperasciititle{Study of BsjpsipipiKK decays} 
\def\papertitle{Beauty meson decays to charmonuim\nobreakdash-like states at LHCb} 
\def\paperkeywords{{High Energy Physics}, {LHCb}} 
\def\papercopyright{\the\year\ CERN for the benefit of the LHCb collaboration} 

\def\paperlicenceurl{https://creativecommons.org/licenses/by/4.0/}

\makeatletter
\g@addto@macro\bfseries{\boldmath}
\makeatother

\usepackage[top=1in, bottom=1.25in, left=1in, right=1in]{geometry}

\usepackage{tikz}

\usepackage{microtype}
\usepackage{lineno}  
\usepackage{xspace} 
\usepackage{caption} 

\usepackage{graphicx}  
\usepackage{color}
\usepackage{colortbl}
\graphicspath{{./figs/}} 
\DeclareGraphicsExtensions{.pdf,.PDF,png,.PNG}

\usepackage{amsmath} 
\usepackage{amssymb}
\usepackage{amsfonts}
\usepackage{upgreek} 

\newcommand*\patchAmsMathEnvironmentForLineno[1]{%
\expandafter\let\csname old#1\expandafter\endcsname\csname #1\endcsname
\expandafter\let\csname oldend#1\expandafter\endcsname\csname
end#1\endcsname
 \renewenvironment{#1}%
   {\linenomath\csname old#1\endcsname}%
   {\csname oldend#1\endcsname\endlinenomath}%
}
\newcommand*\patchBothAmsMathEnvironmentsForLineno[1]{%
  \patchAmsMathEnvironmentForLineno{#1}%
  \patchAmsMathEnvironmentForLineno{#1*}%
}
\AtBeginDocument{%
\patchBothAmsMathEnvironmentsForLineno{equation}%
\patchBothAmsMathEnvironmentsForLineno{align}%
\patchBothAmsMathEnvironmentsForLineno{flalign}%
\patchBothAmsMathEnvironmentsForLineno{alignat}%
\patchBothAmsMathEnvironmentsForLineno{gather}%
\patchBothAmsMathEnvironmentsForLineno{multline}%
\patchBothAmsMathEnvironmentsForLineno{eqnarray}%
}

\usepackage{hyperxmp}

\usepackage[pdftex,
            pdfauthor={\paperauthors},
            pdftitle={\paperasciititle},
            pdfkeywords={\paperkeywords},
            pdfcopyright={Copyright (C) \papercopyright},
            pdflicenseurl={\paperlicenceurl}]{hyperref}

\usepackage[colorinlistoftodos,textsize=scriptsize]{todonotes}

\usepackage[all]{hypcap} 

\usepackage{ifthen} 
\setboolean{uprightparticles}{True} 
\usepackage{xspace} 
\usepackage{upgreek}

\def\lhcb   {\mbox{LHCb}\xspace}

\def\belle  {\mbox{Belle}\xspace}

\def\cdf    {\mbox{CDF}\xspace}

\def\MagUp {\mbox{\em Mag\kern -0.05em Up}\xspace}

\ifthenelse{\boolean{uprightparticles}}%
{

 \def\Pdelta      {\ensuremath{\updelta}\xspace}

 \def\Pmu         {\ensuremath{\upmu}\xspace}

 \def\Ppi         {\ensuremath{\uppi}\xspace}

 \def\Pphi        {\ensuremath{\upphi}\xspace}                 
                  
 \def\Pchi        {\ensuremath{\upchi}\xspace}                 
 \def\Ppsi        {\ensuremath{\uppsi}\xspace}

 \def\PDelta      {\ensuremath{\Delta}\xspace}                 
 \def\PXi         {\ensuremath{\Xi}\xspace}                 
 \def\PLambda     {\ensuremath{\Lambda}\xspace}                 
 \def\PSigma      {\ensuremath{\Sigma}\xspace}                 
 \def\POmega      {\ensuremath{\Omega}\xspace}                 
 \def\PUpsilon    {\ensuremath{\Upsilon}\xspace}

 \def\PB      {\ensuremath{\mathrm{B}}\xspace}                 
                  
 \def\PD      {\ensuremath{\mathrm{D}}\xspace}

 \def\PJ      {\ensuremath{\mathrm{J}}\xspace}                 
 \def\PK      {\ensuremath{\mathrm{K}}\xspace}

 \def\PX      {\ensuremath{\mathrm{X}}\xspace}                 
 \def\PY      {\ensuremath{\mathrm{Y}}\xspace}

 \def\Pc      {\ensuremath{\mathrm{c}}\xspace}

 \def\Pi      {\ensuremath{\mathrm{i}}\xspace}

 \def\Pp      {\ensuremath{\mathrm{p}}\xspace}

 \def\Ps      {\ensuremath{\mathrm{s}}\xspace}

 \def\thebaroffset{0.0em}
}
{

 \def\Pdelta      {\ensuremath{\delta}\xspace}

 \def\Pmu         {\ensuremath{\mu}\xspace}

 \def\Ppi         {\ensuremath{\pi}\xspace}

 \def\Pphi        {\ensuremath{\phi}\xspace}                 
                  
 \def\Pchi        {\ensuremath{\chi}\xspace}                 
 \def\Ppsi        {\ensuremath{\psi}\xspace}                 
                  
 \mathchardef\PDelta="7101
 \mathchardef\PXi="7104
 \mathchardef\PLambda="7103
 \mathchardef\PSigma="7106
 \mathchardef\POmega="710A
 \mathchardef\PUpsilon="7107
                  
 \def\PB      {\ensuremath{B}\xspace}                 
                  
 \def\PD      {\ensuremath{D}\xspace}

 \def\PJ      {\ensuremath{J}\xspace}                 
 \def\PK      {\ensuremath{K}\xspace}

 \def\PX      {\ensuremath{X}\xspace}                 
 \def\PY      {\ensuremath{Y}\xspace}

 \def\Pc      {\ensuremath{c}\xspace}

 \def\Pi      {\ensuremath{i}\xspace}

 \def\Pp      {\ensuremath{p}\xspace}

 \def\Ps      {\ensuremath{s}\xspace}

 \def\thebaroffset{0.18em}
}
\newcommand{\offsetoverline}[2][\thebaroffset]{\kern #1\overline{\kern -#1 #2}}%

\makeatletter
\ifcase \@ptsize \relax
  \newcommand{\miniscule}{\@setfontsize\miniscule{4}{5}}
\or
  \newcommand{\miniscule}{\@setfontsize\miniscule{5}{6}}
\or
  \newcommand{\miniscule}{\@setfontsize\miniscule{5}{6}}
\fi
\makeatother

\DeclareRobustCommand{\optbar}[1]{\shortstack{{\miniscule (\rule[.5ex]{1.25em}{.18mm})}
  \\ [-.7ex] $#1$}}

\def\mup        {{\ensuremath{\Pmu^+}}\xspace}
\def\mun        {{\ensuremath{\Pmu^-}}\xspace} 

\def\squark    {{\ensuremath{\Ps}}\xspace}
\def\squarkbar {{\ensuremath{\overline \squark}}\xspace}

\def\cquark    {{\ensuremath{\Pc}}\xspace}
\def\cquarkbar {{\ensuremath{\overline \cquark}}\xspace}

\def\pion   {{\ensuremath{\Ppi}}\xspace}

\def\pip    {{\ensuremath{\pion^+}}\xspace}
\def\pim    {{\ensuremath{\pion^-}}\xspace}

\def\kaon    {{\ensuremath{\PK}}\xspace}
\def\Kbar    {{\ensuremath{\offsetoverline{\PK}}}\xspace}

\def\KorKbar {\kern \thebaroffset\optbar{\kern -\thebaroffset \PK}{}\xspace}

\def\Kp      {{\ensuremath{\kaon^+}}\xspace}
\def\Km      {{\ensuremath{\kaon^-}}\xspace}

\def\Kstarz  {{\ensuremath{\kaon^{*0}}}\xspace}
\def\Kstarzb {{\ensuremath{\Kbar{}^{*0}}}\xspace}

\def\D       {{\ensuremath{\PD}}\xspace}

\def\DorDbar {\kern \thebaroffset\optbar{\kern -\thebaroffset \PD}\xspace}

\def\Ds      {{\ensuremath{\D^+_\squark}}\xspace}

\def\B       {{\ensuremath{\PB}}\xspace}

\def\BorBbar {\kern \thebaroffset\optbar{\kern -\thebaroffset \PB}\xspace}

\def\Bd      {{\ensuremath{\B^0}}\xspace}

\def\BdorBdbar {\kern \thebaroffset\optbar{\kern -\thebaroffset \Bd}\xspace}
\def\Bu      {{\ensuremath{\B^+}}\xspace}

\def\Bp      {{\ensuremath{\Bu}}\xspace}

\def\Bs      {{\ensuremath{\B^0_\squark}}\xspace}

\def\BsorBsbar {\kern \thebaroffset\optbar{\kern -\thebaroffset \Bs}\xspace}

\def\jpsi     {{\ensuremath{{\PJ\mskip -3mu/\mskip -2mu\Ppsi\mskip 2mu}}}\xspace}
\def\psitwos  {{\ensuremath{\Ppsi{(\rm{2S})}}}\xspace}

\def\chicone  {{\ensuremath{\Pchi_{\cquark 1}}}\xspace}
\def\chiconex {{\ensuremath{\Pchi_{\cquark 1}(3872)}}\xspace}

\def\chictwo  {{\ensuremath{\Pchi_{\cquark 2}}}\xspace}

\def\Y#1S{\ensuremath{\PUpsilon{(#1S)}}\xspace}

\def\proton      {{\ensuremath{\Pp}}\xspace}
\def\antiproton  {{\ensuremath{\overline \proton}}\xspace}

\def\LorLbar     {\kern \thebaroffset\optbar{\kern -\thebaroffset \PLambda}\xspace}

\def\BF         {{\ensuremath{\mathcal{B}}}\xspace}
\def\BRN         {\BF}

\newcommand{\decay}[2]{\ensuremath{#1\!\to #2}\xspace} 

\def\to                 {\ensuremath{\rightarrow}\xspace}

\def\JpsiPiPi     {{\jpsi\pip\pim}~}

\def\AT#1     {\ensuremath{A_{\mathrm{T}}^{#1}}\xspace}           

\def\C#1      {\ensuremath{\mathcal{C}_{#1}}\xspace}                       
\def\Cp#1     {\ensuremath{\mathcal{C}_{#1}^{'}}\xspace}                    
\def\Ceff#1   {\ensuremath{\mathcal{C}_{#1}^{\mathrm{(eff)}}}\xspace}        
\def\Cpeff#1  {\ensuremath{\mathcal{C}_{#1}^{'\mathrm{(eff)}}}\xspace}       
\def\Ope#1    {\ensuremath{\mathcal{O}_{#1}}\xspace}                       
\def\Opep#1   {\ensuremath{\mathcal{O}_{#1}^{'}}\xspace}                    

\newcommand{\aunit}[1]{\ensuremath{\text{\,#1}}}       

\newcommand{\tev}{\aunit{Te\kern -0.1em V}\xspace}
\newcommand{\gev}{\aunit{Ge\kern -0.1em V}\xspace}
\newcommand{\mev}{\aunit{Me\kern -0.1em V}\xspace}
\newcommand{\kev}{\aunit{ke\kern -0.1em V}\xspace}
\newcommand{\ev}{\aunit{e\kern -0.1em V}\xspace}
\newcommand{\mevc}{\ensuremath{\aunit{Me\kern -0.1em V\!/}c}\xspace}
\newcommand{\gevc}{\ensuremath{\aunit{Ge\kern -0.1em V\!/}c}\xspace}
\newcommand{\mevcc}{\ensuremath{\aunit{Me\kern -0.1em V\!/}c^2}\xspace}
\newcommand{\gevcc}{\ensuremath{\aunit{Ge\kern -0.1em V\!/}c^2}\xspace}

\def\fb   {\ensuremath{\aunit{fb}}\xspace}
\def\invfb   {\ensuremath{\fb^{-1}}\xspace}

\def\gsim{{~\raise.15em\hbox{$>$}\kern-.85em
          \lower.35em\hbox{$\sim$}~}\xspace}
\def\lsim{{~\raise.15em\hbox{$<$}\kern-.85em
          \lower.35em\hbox{$\sim$}~}\xspace}

\def\sqs   {\ensuremath{\protect\sqrt{s}}\xspace}

\def\tell1  {TELL1\xspace}
\def\ukl1   {UKL1\xspace}

\newcommand{\kevcc}{\ensuremath{\aunit{ke\kern -0.1em V\!/}c^2}\xspace}

\def\chiconex  {\ensuremath{\chicone(3872)}\xspace}

\usepackage{boldline}
\usepackage{bm}

\usepackage{cite} 
\usepackage{mciteplus}

\usepackage{longtable} 

\newcolumntype{d}[1]{D{,}{\,\pm\,}{#1} }
\newcolumntype{f}[1]{D{,}{.}{#1} }

\usepackage{tikz}

\begin{document}

\renewcommand{\thefootnote}{\fnsymbol{footnote}}
\setcounter{footnote}{1}


\begin{titlepage}
\pagenumbering{roman}

\vspace*{-1.5cm}
\vspace*{1.5cm}
\noindent
\begin{tabular*}{\linewidth}{lc@{\extracolsep{\fill}}r@{\extracolsep{0pt}}}
\\
 & & \today \\ 
\end{tabular*}

\vspace*{3.2cm}

{\normalfont\bfseries\boldmath\huge
\begin{center}
  \papertitle 
\end{center}
}

\vspace*{0.2cm}

\begin{center}
Tatiana Ovsiannikova\footnote{\tt{E-mail:}\href{email:Tatiana.Ovsiakkinova@cern.ch}{\tt{Tatiana.Ovsiannikova@cern.ch}}} on behalf of the~LHCb collaboration 
\\{\it Institute for Theoretical and Experimental Physics, NRC Kurchatov Institute, B.~Cheremushkinskaya~25, Moscow, 117218, Russia.}

\end{center}

\vspace{2cm}

\begin{abstract}

The decays $\decay{\Bs}{\jpsi\pip\pim\Kp\Km}$ are studied
using  a~data set corresponding 
to an~integrated luminosity of 9\invfb, 
collected with the~LHCb detector 
in proton\nobreakdash-proton collisions
at centre\nobreakdash-of\nobreakdash-mass 
energies of 7, 8 and 13\tev.
The~decays
\mbox{$\decay{\Bs}{\jpsi\Kstarz\Kstarzb}$}
and 
\mbox{$\decay{\Bs}
{\chicone(3872)}\Kp\Km$},
where the~$\Kp\Km$~pair does not originate from 
a~\Pphi~meson, 
are 
observed for the~first time. 
Precise measurements of the~ratios of branching fractions 
between intermediate 
$\chicone(3872)\Pphi$,
$\jpsi\Kstarz\Kstarzb$,
$\psitwos\Pphi$ and 
$\chicone(3872)\Kp\Km$~states
are  reported.
A~structure, denoted
as $\PX(4740)$,  
is observed  in the~$\jpsi\Pphi$~mass spectrum
with a significance in excess of 5.3 standard deviation.
In~addition, 
the~most precise single measurement of  
the~mass of the~\Bs~meson is performed.
\end{abstract}

\vspace{\fill}

\begin{center}
  Talk given at 23th International Conference in Quantum Chromodynamics (QCD 20,  35th anniversary),  27  - 30 November 2020, Montpellier - FR
\end{center}



\end{titlepage}


\newpage
\setcounter{page}{2}
\mbox{~}
%

\cleardoublepage


\renewcommand{\thefootnote}{\arabic{footnote}}
\setcounter{footnote}{0}



\pagestyle{plain} 
\setcounter{page}{1}
\pagenumbering{arabic}


%

\section{Introduction}
Decays of beauty hadrons to final states with charmonia
provide a~unique laboratory to study 
the~properties of charmonia and 
charmonium-like states. 
A~plethora of new charmonium-like states has been 
observed in the~decays of beauty mesons,  
such as the~$\chicone(3872)$ meson~\cite{Xbelle}
and numerous tetraquark candidates~\cite{Choi:2007wga,
Mizuk:2009da,
Chilikin:2013tch,
LHCb-PAPER-2014-014,
LHCb-PAPER-2015-038,
LHCb-PAPER-2016-015,
LHCb-PAPER-2016-018,
LHCb-PAPER-2016-019,
LHCb-PAPER-2018-034,
LHCb-PAPER-2018-043}. 
The~nature of many exotic charmonium-like  candidates 
remains unclear. 
A~comparison of production rates 
with respect to those  
of conventional  charmonium states 
in decays 
of beauty hadrons can shed light on 
their~production mechanisms. 

The reported results  
are based on the data samples 
collected by the LHCb experiment in proton-proton\,(\proton\proton) 
collisions at centre-of-mass energies $\sqs= 7,8$  and $13\tev$ between 2011 and 2018. 

\section{Study of the 
\mbox{$\decay{\Bp }{ \jpsi \pip \pim \Kp}$} decays.}
Candidate \mbox{$\decay{\Bp }{ \jpsi \pip \pim \Kp}$} decays are reconstructed 
using the~$\jpsi \to \mup \mun$ decay mode. 
A~loose pre-selection 
is applied, followed by 
a~multivariate classifier based on 
a~decision tree with gradient boosting~\cite{LHCb-PAPER-2020-009}.  
\begin{figure}[htb]
	\setlength{\unitlength}{1mm}
	\centering
	\begin{picture}(160,65)
	\definecolor{root8}{rgb}{0.35, 0.83, 0.33}
		\put(0,0){\includegraphics*[width=80mm,height=65mm]{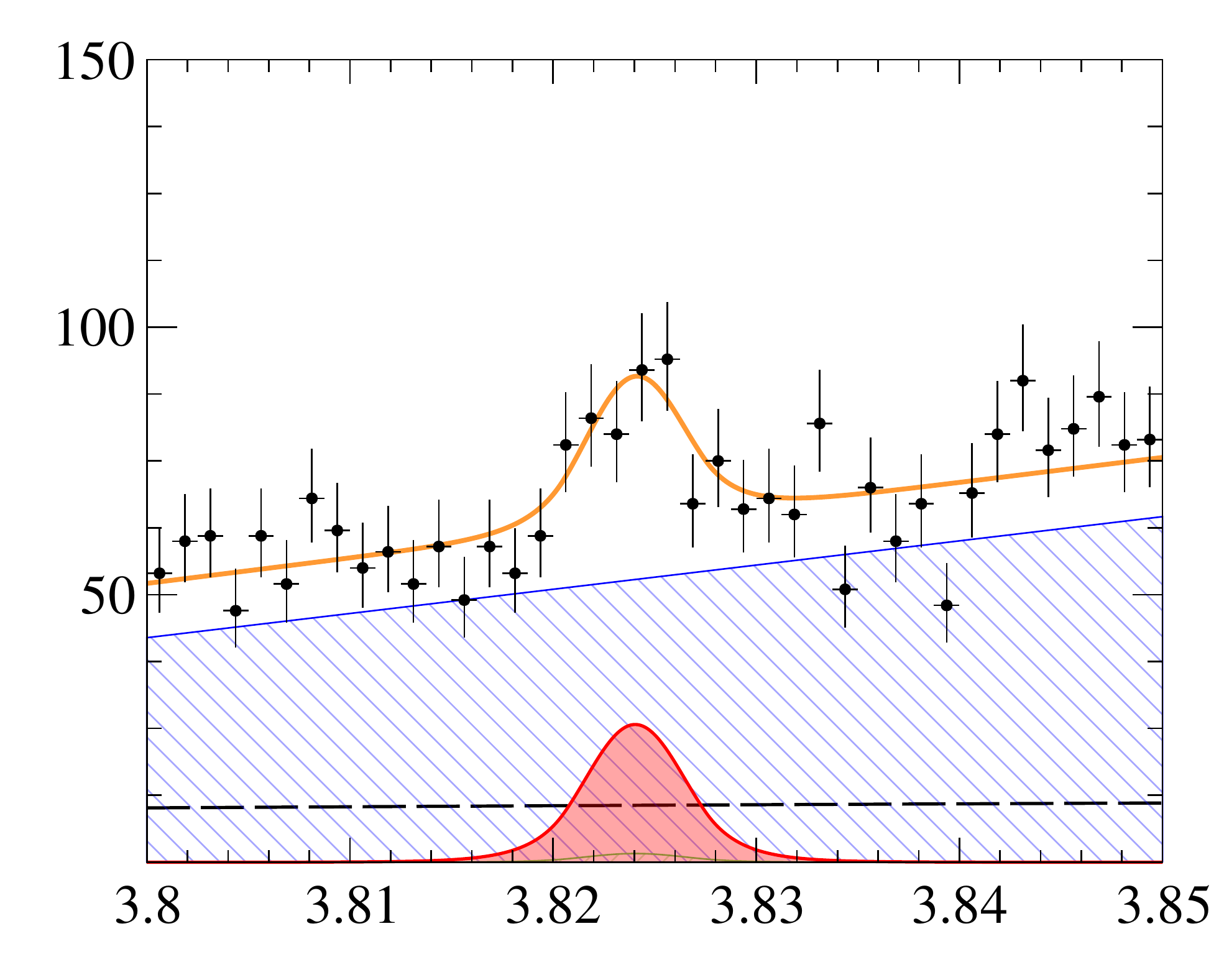}}
    \put(80,0){\includegraphics*[width=80mm,height=65mm]{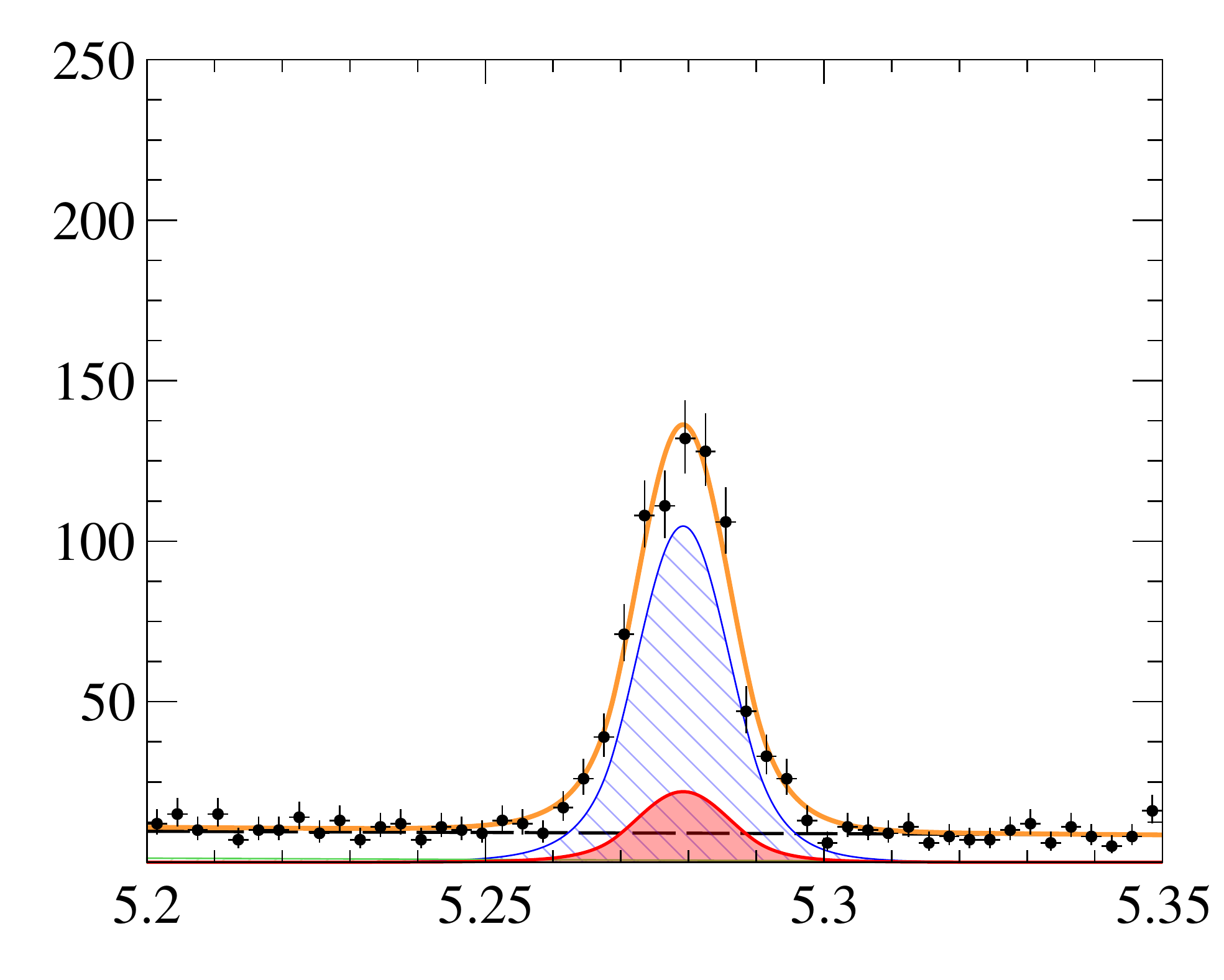}}

	\put(  -2,22){\begin{sideways}\small{Candidates/$(3\mevcc)$}\end{sideways}}
	
	\put(  78,18){\begin{sideways}\small{Candidates/$(1.25\mevcc)$}\end{sideways}}
	
	\put(30 ,0){$m_{\jpsi\pip\pim\Kp}$}

	\put( 65,-1){$\left[\!\gevcc\right]$}

	\put(120 ,0){$m_{\jpsi\pip\pim}$}

	\put( 145,-1){$\left[\!\gevcc\right]$}

	\put( 63,55){\small\lhcb}
    \put( 143,55){\small\lhcb}

    \put(12,55.5) {\begin{tikzpicture}[x=1mm,y=1mm]\filldraw[fill=red!35!white,draw=red,thick]  (0,0) rectangle (8,3);\end{tikzpicture} }
    \put(12,51.5){\begin{tikzpicture}[x=1mm,y=1mm]\draw[thin,blue,pattern=north west lines, pattern color=blue]  (0,0) rectangle (8,3);\end{tikzpicture} }
    \put(12,47.5){\begin{tikzpicture}[x=1mm,y=1mm]\draw[thin,root8,pattern=north east lines, pattern color=root8]  (0,0) rectangle (8,3);\end{tikzpicture} }
	\put(12,44.5){\color[RGB]{0,0,0}     {\hdashrule[0.0ex][x]{8mm}{1.0pt}{2.0mm 0.3mm} } }
	\put(12,40.5){\color[RGB]{255,153,51} {\rule{8mm}{2.0pt}}}
	
	\put( 22,56.3){\scriptsize{\decay{\Bp}{\PX_{\cquark\cquarkbar}\Kp}}}
	\put( 22,53){\scriptsize{\decay{\Bp}{\left(\jpsi\pip\pim\right)_{\mathrm{NR}}\Kp}}}
	\put( 22,49.5){\scriptsize{comb. $\PX_{\cquark\cquarkbar}\Kp$}}
	\put( 22,44){\scriptsize{comb. bkg.}}
	\put( 22,40){\scriptsize{total}}
	\end{picture}
	\caption {\small 
	Distributions of 
	the~(left)\,\jpsi\pip\pim\Kp and 
	(right)\,\JpsiPiPi mass
	for the~selected 
	\mbox{$\decay{\Bp}{\Ppsi_{2}(3823)  \Kp}$}~candidates\,(points with error bars)~\cite{LHCb-PAPER-2020-009,CERN-THESIS-2020-204}.}
	\label{fig:signal_fit1}
\end{figure}

The yields for the~\mbox{$\Bp \to \jpsi \pip \pim \Kp$}
decays via intermediate  
\mbox{$\decay{\chicone(3872)}{\jpsi\pip\pim}$},
\mbox{$\decay{\Ppsi_2(3823)}{\jpsi\pip\pim}$} and 
\mbox{$\decay{\psitwos}{\jpsi\pip\pim}$}~chains
are determined using a simultaneous unbinned extended
maximum-likelihood fit to the $ \jpsi \pip \pim\Kp$ 
mass and the $ \jpsi \pip \pim$  mass distributions. 
The fit is performed in the three non-overlapping regions 
around the $\Ppsi_{2}(3823)$, $\chicone(3872)$ and $\psitwos$ masses.
To improve the resolution on the $ \jpsi \pip \pim$ mass 
and to eliminate a small correlation between 
$m_{\jpsi \pip \pim \Kp}$ and  $m_{\jpsi \pip \pim}$ variables, 
the  $m_{\jpsi \pip \pim}$ variable is computed using 
a kinematic fit  
that constrains 
the mass of the $\Bp$ candidate to its known value~\cite{PDG2020}. 
The signal yields are determined to be $137\pm 26$ 
events for the \mbox{$\decay{\Bp }{  \Ppsi_{2}(3823)  \Kp}$} decay  
which correspond to statistical significance above 5.1$\sigma$.  
The fit to the mass distribution for the signal channel 
are shown in figure~\ref{fig:signal_fit1}. 
Also the significant signal yield is observed for 
the \mbox{$\decay{\Bp }{  \chicone(3872)  \Kp}$} decay $4230\pm 70$ 
events which allows 
the precise determination of the~parameters 
of the~$\chicone(3872)$ state, in particular
for the first time the non-zero width for 
the $\chicone(3872)$ state is observed with significance more than 
5~standard deviations
\begin{linenomath}
\begin{equation*}
\Gamma_{\chicone(3872)}=0.96^{+0.19}_{-0.18} \pm 0.21\mev\, ,
\end{equation*}
\end{linenomath}
where the first uncertainty is statistical and the second is systematic. 
The~value of the~Breit--Wigner width
agrees well with the~value from the~analysis of 
a~large sample of \mbox{$\decay{\chiconex}{\jpsi\pip\pim}$}
decays from the~inclusive decays
of beauty hadrons~\cite{LHCb-PAPER-2020-008}.
The~improved upper limit for the width of 
the  $\Ppsi_{2}(3823)$ meson is found to be 
\mbox{$\Gamma_{\Ppsi_{2}(3823)}<5.2\,(6.6)\mev$} for $90\,(95)\%$~C.L.
The~mass differences 
between $\Ppsi_{2}(3823)$, $\chicone(3872)$  and $\psitwos$ mesons, 
\mbox{$\Pdelta m^{\PX}_{\PY}\equiv m_{\PX}-m_{\PY}$},
are 
measured to be 
\begin{linenomath}
\begingroup
\allowdisplaybreaks
\begin{eqnarray*}
\Pdelta m^{\chicone(3872)}_{{\Ppsi_{2}(3823)}} & = &  
  \phantom{0}47.50\pm 0.53 \pm 0.13 \mevcc\,,\\ 
\Pdelta m^{{\Ppsi_{2}(3823)}}_{{\psitwos}} & = & 
137.98\pm 0.53 \pm 0.14 \mevcc\,, \\ 
\Pdelta m^{{\Ppsi_{2}(3823)}}_{{\psitwos}} & = & 
185.49\pm 0.06 \pm 0.03 \mevcc\,,
\end{eqnarray*}
\endgroup
\end{linenomath}
where the~first uncertainty is statistical and the second is systematic. 
Using the~measured mass difference the~binding energy 
of the~$\chicone(3872)$ state is calculated to be 
\mbox{$\delta E=0.12\pm 0.13 \mev$}.  It is consistent 
with zero within uncertainties, that are currently 
dominated   by the~uncertainty  
for the~charged and neutral kaon mass measurements. 

The measured yields of 
the~\mbox{$\decay{\Bp}{\chicone(3872) \Kp}$},
\mbox{$\decay{\Bp}{\Ppsi_{2}(3823) \Kp}$} and 
\mbox{$\decay{\Bp}{\psitwos \Kp}$} signal decays
allow for a precise determination of the ratios of 
the~branching fractions:
\begingroup
\allowdisplaybreaks
\begin{eqnarray*}
\dfrac{ \BRN_{\decay{\Bp }{  \Ppsi_{2}(3823)  \Kp}} \times \BRN_{\decay{\Ppsi_{2}(3823)} {\jpsi \pip \pim }}}
{\BRN_{\decay{\Bp }{  \chicone(3872)  \Kp}} \times 
\BRN_{\decay{\chicone(3872) } { \jpsi \pip \pim}  } }
& = &     \left( 3.56 \pm 0.67 \pm 0.11 \right) \times 10^{-2} \,, 
\\
\dfrac{ \BRN_{\decay{\Bp }{ \Ppsi_{2}(3823)  \Kp }} \times 
\BRN_{\left( \decay{\Ppsi_{2}(3823)} {\jpsi \pip \pim }\right)}}
{\BRN_{\decay{\Bp }{  \psitwos \Kp} } \times \BRN_{ \decay{\psitwos} {  \jpsi \pip \pim }}}
& = &  (1.31 \pm 0.25 \pm 0.04) \times 10^{-3} \,,  \\
\dfrac{ \BRN_{\decay{\Bp }{  \chicone(3872)  \Kp }} \times \BRN_{\decay{\chicone(3872) } { \jpsi \pip \pim} }}
{\BRN_{\decay{\Bp }{ \psitwos\Kp} } \times \BRN_{ \decay{\psitwos} { \jpsi \pip \pim} }} 
& = & 
(3.69 \pm 0.07  \pm 0.06) \times 10^{-2} \,.
\end{eqnarray*}
\endgroup

\section{Study of the~\mbox{$\decay{\Bs}{ \jpsi \pip \pim\Kp\Km}$} decays}
The \mbox{$\decay{\Bs}{ \jpsi \pip \pim  \Kp \Km}$} decays 
are reconstructed using  selection criteria
 based on kinematics, particle identification and topology~\cite{LHCb-PAPER-2020-035}.
The yields of 
\mbox{$\decay{\Bs}{ \jpsi \pip \pim \Kp \Km}$}  decays 
via intermediate 
\mbox{$\decay{\psitwos}{ \jpsi \pip \pim}$}  and 
\mbox{$\decay{\chicone(3872) }{ \jpsi \pip \pim}$} chains
are determined using a three-dimensional unbinned 
extended maximum-likelihood fit to the 
$ \jpsi \pip \pim  \Kp \Km$, 
$\jpsi \pip \pim $ and $\Kp \Km$ mass distributions. 
The fit  is performed simultaneously
in two separate regions corresponding to $\decay{\Bs}{ \chicone(3872)\Pphi}$ and \mbox{$\decay{\Bs}{  \psitwos\Pphi}$}
signals as described above. 

\begin{figure*}[htb]
	\setlength{\unitlength}{1mm}
	\centering
	\begin{picture}(160,65)
	\definecolor{root8}{rgb}{0.35, 0.83, 0.33}

    \put(0,0){\includegraphics*[width=80mm,height=65mm]{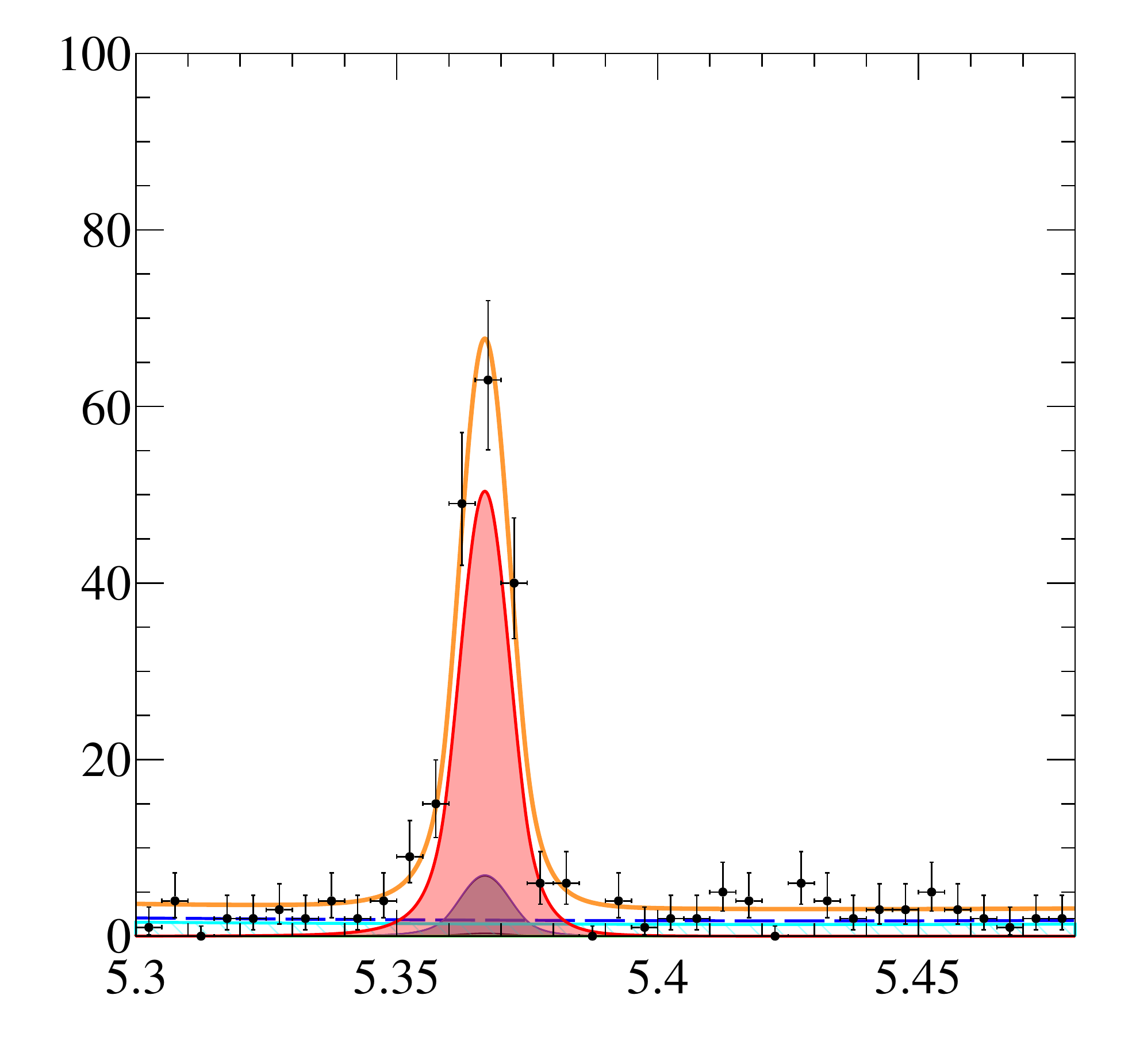}}
		\put(80,0){\includegraphics*[width=80mm,height=65mm]{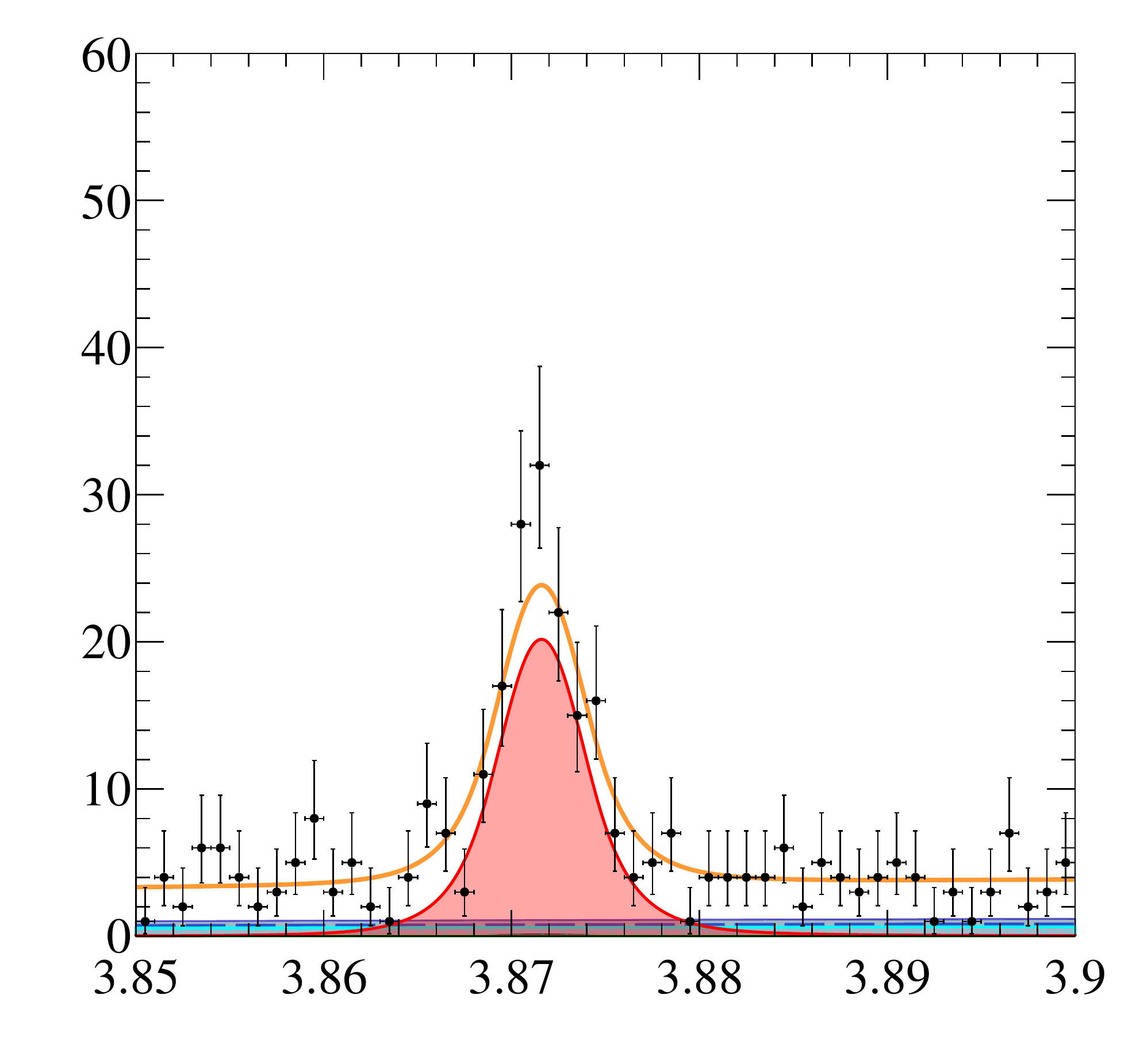}}

	\put(  -2,22){\begin{sideways}\small{Candidates/$(5\mevcc)$}\end{sideways}}
	
	\put(  80,22){\begin{sideways}\small{Candidates/$(1\mevcc)$}\end{sideways}}

	\put(30 ,0){$m_{\jpsi\pip\pim\Kp \Kp}$}
	
	\put(120 ,0){$m_{\jpsi\pip\pim}$}

    \put( 65,-1){$\left[\!\gevcc\right]$}

	\put( 145,-1){$\left[\!\gevcc\right]$}

	\put( 15,50){\scriptsize$3.864<m_{\jpsi\pip\pim}<3.880\gevcc$}
	\put( 15,56){\scriptsize$1.01<m_{\Kp\Km}<1.03\gevcc$}
	\put( 95,50){\scriptsize$5.350<m_{\jpsi\pip\pim\Kp\Km}<5.384\gevcc$}
	\put( 95,56){\scriptsize$1.01<m_{\Kp\Km}<1.03\gevcc$}
	\put( 63,55){\small\lhcb}

	\put( 143,55){\small\lhcb}

	\end{picture}
		\caption {\small 
	Distributions of 
	the~(left)\,\jpsi\pip\pim\Kp \Km and 
	(right)\,\jpsi \pip \pim mass
	for selected \mbox{$\decay{\Bs}{ \chicone(3872)  \Pphi}$}~candidates\,(points with error bars)~\cite{LHCb-PAPER-2020-035}. The red filled area corresponds to the \mbox{$\decay{\Bs}{  \chicone(3872)  \Pphi}$} signal. The orange line is the total fit. }
	\label{fig:signal_fit3}
\end{figure*}


The observed signal yield for the \mbox{$\decay{\Bs}{ \chicone(3872)\Pphi}$} decays is found to be $154\pm 15$ events which corresponds to the statistical significance more than $10 \sigma$ deviation.  The fit to the mass distribution for the signal channel are shown in figure ~\ref{fig:signal_fit3}. 
Using the obtained signal yields for \mbox{$\decay{\Bs}{  \chicone(3872)\Pphi}$} and \mbox{$\decay{\Bs}{ \psitwos\Pphi}$} channels and corresponding efficiency ratio the following branching fraction is calculated:
\begin{linenomath}
\begin{eqnarray*}
\dfrac{ \BRN_{\decay{\Bs}{  \chicone(3872)  \Pphi}} \times \BRN_{ \decay{\chicone(3872) } { \jpsi \pip \pim }}}
{\BRN_{\decay{\Bs}{  \psitwos\Pphi }} \times \BRN_{\decay{\psitwos }{ \jpsi \pip \pim }}}
& = & (2.42 \pm 0.23  \pm 0.07) \times 10^{-2}\,.
\end{eqnarray*}
\end{linenomath}
The obtained value is found to be in a good  agreement with 
the~recent result  by the~CMS collaboration~\cite{Sirunyan:2020qir} but is more precise.

\begin{figure}[htb]
	\setlength{\unitlength}{1mm}
	\centering
	\begin{picture}(150,120)
	\definecolor{root8}{rgb}{0.35, 0.83, 0.33}

    \put(0,0){\includegraphics*[width=150mm,height=120mm]{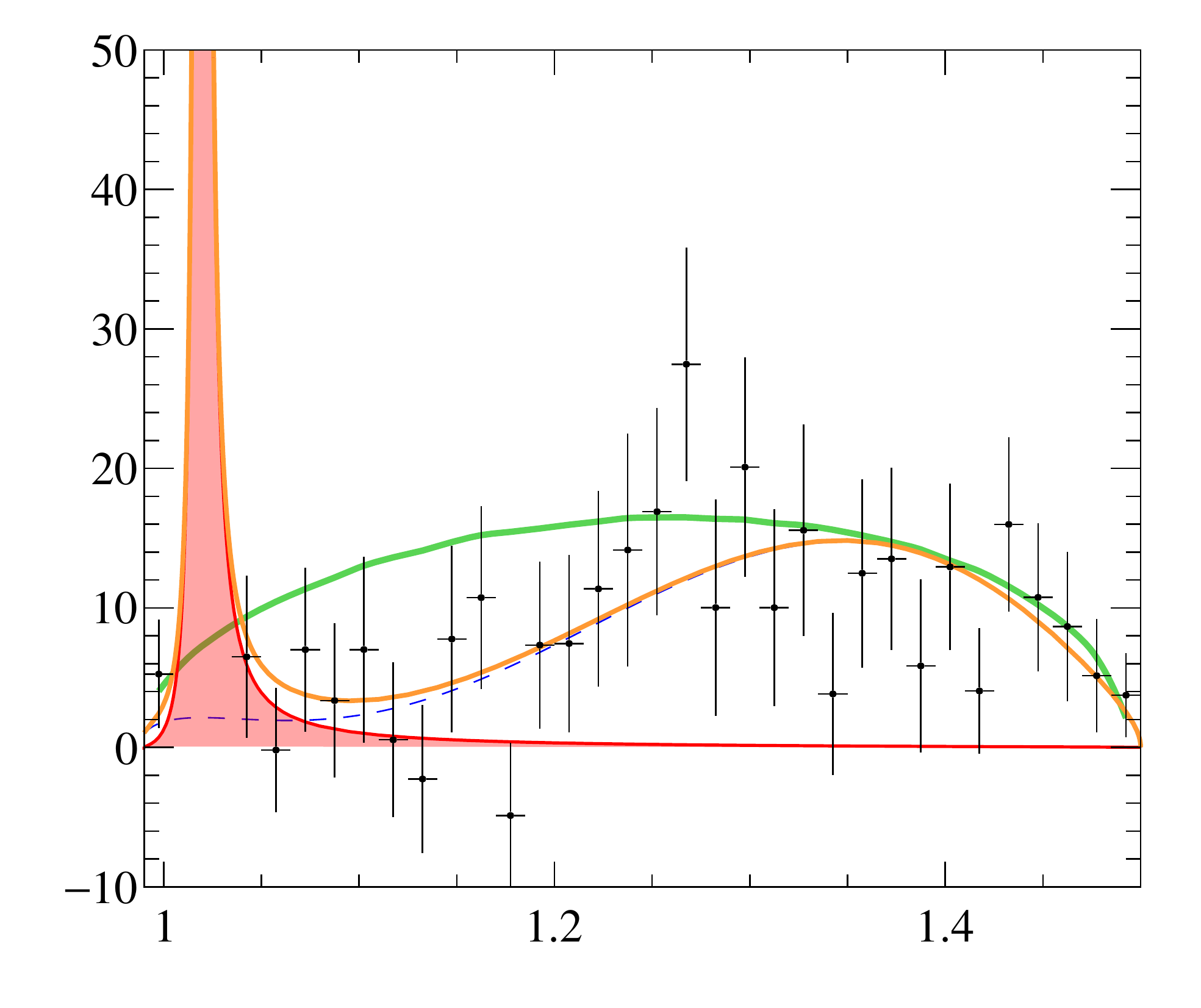}}

	\put(0,  70){\begin{sideways}\Large{Yields/$(15\mevcc)$}\end{sideways}}

	 \put(75, 0) {\Large$m_{\Kp \Kp}$}

	\put(125, 0){\Large$\left[\!\gevcc\right]$}

   \put(40,100) {\begin{tikzpicture}[x=1mm,y=1mm]\filldraw[fill=red!35!white,draw=red,thick]  (0,0) rectangle (8,3);\end{tikzpicture} }
	\put(40,94){\color[RGB]{85,83,246}     {\hdashrule[0.0ex][x]{8mm}{1.0pt}{1.0mm 0.5mm} } }
	\put(40,88){\color[RGB]{255,153,51} {\rule{8mm}{1.0pt}}}
	\put(55,100){ $\decay{\Bs}{\chicone(3872)\Pphi}$}
	\put( 55,94){ \decay{\Bs}{\chicone(3872)\Kp \Km}}
	\put( 55,88){ total}

	\put(120,103){\Large\lhcb}

	\end{picture}
	\caption {\small 
	 Background-subtracted $\Kp\Km$~mass distribution 
	for selected \mbox{$\decay{\Bs}{  \chicone(3872)  \Kp \Km}$}~candidates\,(points with error bars)~\cite{LHCb-PAPER-2020-035}. 
	A~fit, described in the~text, is overlaid. 
	The~red filled area corresponds to the \mbox{$\decay{\Bs}{  \chicone(3872)  \Pphi}$} signal. 
		The~blue dashed line corresponds to the \mbox{$\decay{\Bs}{  \chicone(3872)  \Kp \Km}$} component. 
	The~orange line is the total fit. 
	The~expectation for phase-space simulated decays is shown as a~green solid line. }
	\label{fig:signal_fit5}
\end{figure}

The decay \mbox{$\decay{\Bs}{  \chicone(3872)\Kp \Km}$}  
where the $\Kp \Km$ pair does not originate from
a $\Pphi$ meson, is studied using a sample of selected 
\mbox{$\decay{\Bs}{\jpsi\pip\pim\Kp\Km}$} signal decays.
A two-dimensional unbinned extended maximum-likelihood fit 
is performed to the~$\jpsi \pip \pim$ and $\jpsi \pip \pim \Kp \Km$ mass distributions. 
The yield of 
the~\mbox{$\decay{\Bs}{  \chicone(3872)\Kp \Km}$} signal decays 
is \mbox{$378 \pm 33$},
that is significantly large than 
the~yield of the~\mbox{$\decay{\Bs}{ \chicone(3872)\Pphi}$}~decays, 
indicating a large \mbox{$\decay{\Bs}{  \chicone(3872)\Kp \Km}$}~contribution. 
The background-subtracted and 
efficiency-corrected $\Kp\Km$~mass distribution 
of the \mbox{$\decay{\Bs}{ \chicone(3872)\Kp \Km}$} candidates is shown in figure~\ref{fig:signal_fit5}. 
The $\Kp \Km$ mass
distribution for $m_{\Kp \Km}>1.1~\gevcc$ region 
cannot be described by phase-space shape, 
and possibly contains contributions from 
the $f_{0}(980)$, $f_{2}(1270)$, $f_{0}(1370)$ and $f^{'}_{2}(1520)$  
resonances decaying
to a pair of kaons, as has been observed  in 
\mbox{$\decay{\Bs}{  \jpsi \Kp \Km}$} decays~\cite{LHCb-PAPER-2012-040,LHCb-PAPER-2017-008}. Therefore a component that accounts for 
     non\nobreakdash-resonant
     \mbox{$\decay{\Bs}{\chicone(3872)\Kp\Km}$}~decays  
     and decays via broad~high\nobreakdash-mass
     $\Kp\Km$~intermediate states, modelled 
     by a~product of a~phase\nobreakdash-space function
     for three\nobreakdash-body
     \mbox{$\decay{\Bs}{\chicone(3872)\Kp\Km}$}~decays and 
     a~third\nobreakdash-order  polynomial function. 
An~amplitude analysis of a larger data sample would be required to 
properly disentangle individual contributions. 
However, a~narrow $\Pphi$ component can be separated from the non-$\Pphi$ components
using an unbinned maximum-likelihood fit to the background-subtracted and
efficiency-corrected $\Kp \Km$ mass distribution. 
The~fraction of
the~\mbox{$\decay{\Bs}{  \chicone(3872)\Kp \Km}$} signal component is found 
to be $(38.9 \pm 4.9)\%$ and further propagated to the branching fraction ratio:
\begin{linenomath}
\begin{eqnarray*}
\dfrac{ \BRN_{\decay{\Bs}{ \chicone(3872)  (\Kp \Km)_{\text{non-}\Pphi}} }}
{\BRN_{\decay{\Bs}{ \chicone(3872) \Pphi} } 
\times \BRN_{ \decay{\Pphi}{  \Kp\Km}  }}= 1.57 \pm 0.32  \pm 0.12 \,.
\end{eqnarray*}
\end{linenomath}


The yield of \mbox{$\decay{\Bs}{ \jpsi\Kstarz\Kstarzb}$} decays 
is determined using a three-dimensional unbinned extended maximum-likelihood fit 
to the~$\jpsi \pip \pim \Kp \Km$, $ \Km \pip$ and $\Kp \pim$ mass distributions. 
Using the~obtained signal yields and 
the~corresponding efficiency ratio the branching fraction ratio is calculated:
\begin{eqnarray*}
 \dfrac{\BRN_{\decay{\Bs}{ \jpsi\Kstarz\Kstarzb}} 
 \times \BRN_{\decay{\Kstarz} { \Kp \pim }}^2}{\BRN_{\decay{\Bs}{ \psitwos \Pphi}} \times 
 \BRN_{ \decay{\psitwos }{ \jpsi \pip \pim} }
 \times \BRN_{\decay{\Pphi}{  \Kp\Km} }} & =  & 1.22 \pm 0.03 \pm 0.04\,.
\end{eqnarray*}

\begin{figure}[htb]
	\setlength{\unitlength}{1mm}
	\centering
	\begin{picture}(150,120)
	\definecolor{root8}{rgb}{0.35, 0.83, 0.33}
		\put(0,0){\includegraphics*[width=150mm,height=120mm]{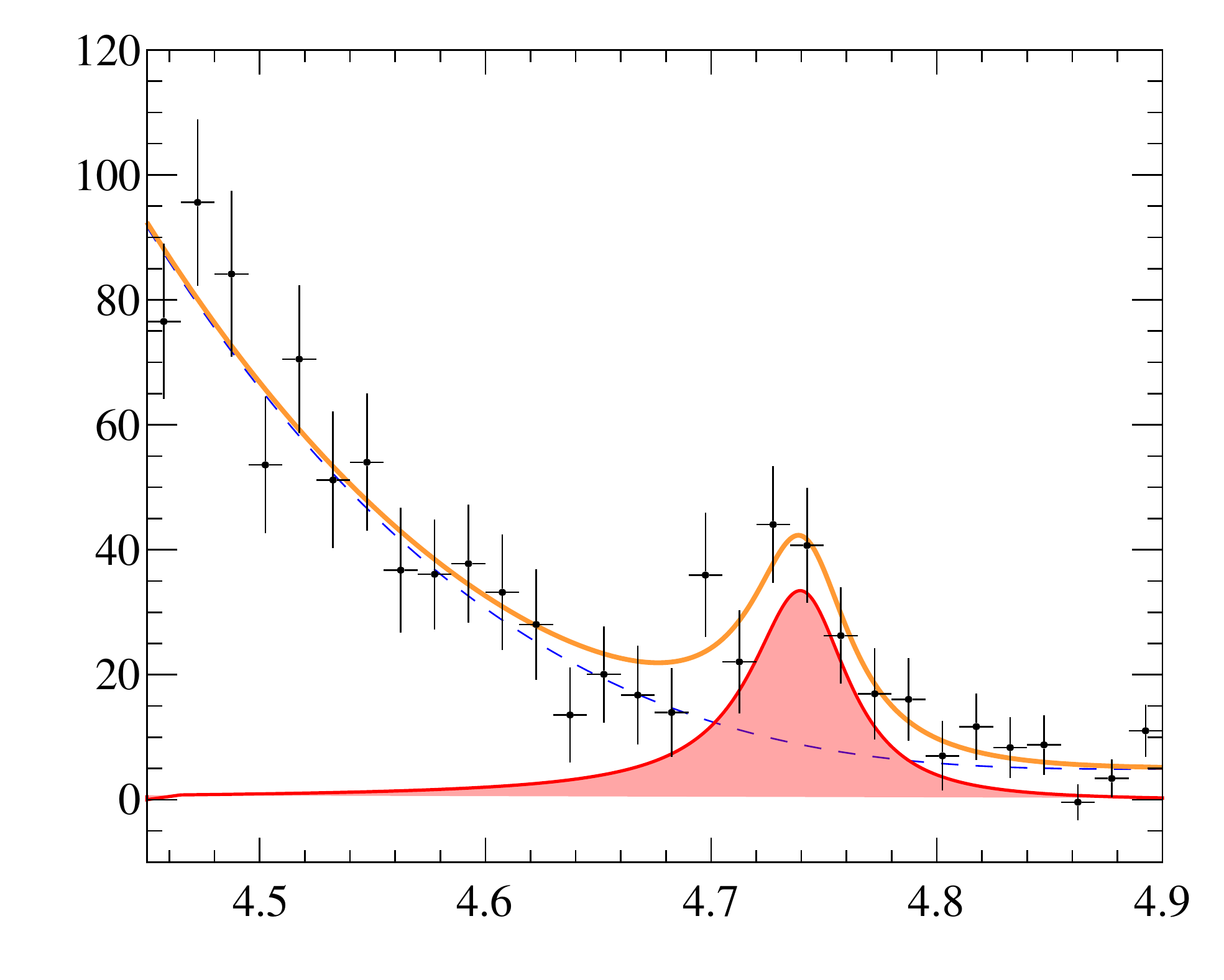}}
    \put( 75,0){\Large$m_{\jpsi\Pphi}$}
	\put(125,0){\Large$\left[\!\gevcc\right]$}

    \put(0,70){\begin{sideways}\Large{Yields/$(15\mevcc)$}\end{sideways}}
	\put(120,102){\Large\lhcb} 

   \put(40,100){\begin{tikzpicture}[x=1mm,y=1mm]\filldraw[fill=red!35!white,draw=red,thick]  (0,0) rectangle (8,3);\end{tikzpicture} }
	\put(40,94){\color[RGB]{85,83,246}     {\hdashrule[0.0ex][x]{8mm}{1.0pt}{1.0mm 0.5mm} } }
	\put(40,88){\color[RGB]{255,153,51} {\rule{8mm}{1.0pt}}}
	\put(55,100){ $\decay{\Bs}{\PX(4740)\pip\pim}$}
	\put( 55,94){ \decay{\Bs}{\jpsi \pip \pim \Pphi}}
	\put( 55,88){ total}

		\end{picture}
		\caption{\small Background-subtracted $\jpsi\Pphi$~mass distribution
for selected \mbox{$\decay{\Bs}{  \jpsi \pip \pim \Pphi}$}~candidates\,(points 
with error bars)~\cite{LHCb-PAPER-2020-035}.  
The red filled area corresponds to 
the \mbox{$\decay{\Bs}{  X(4740) \pip \pim}$} signal. The orange line is the total fit.}
		\label{fig:signal_fit6}
\end{figure}
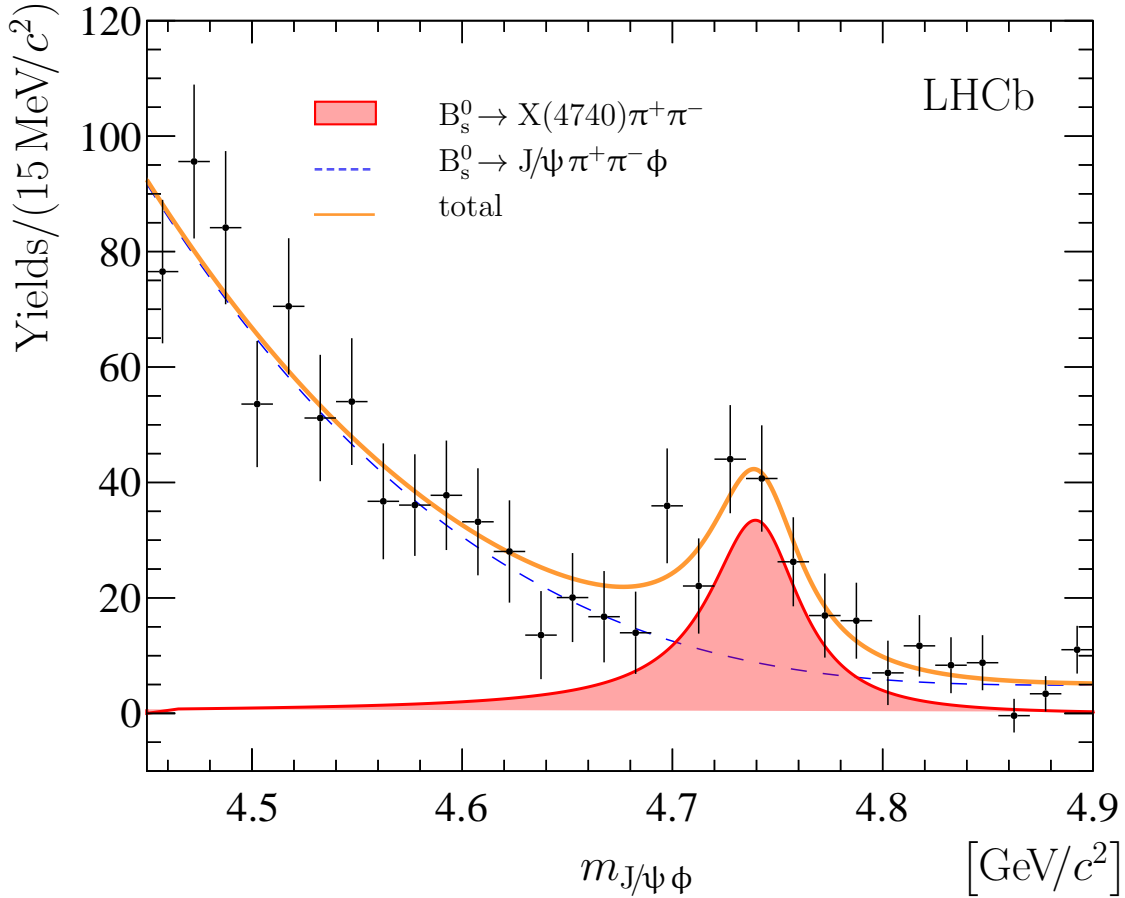

The~$\jpsi\Pphi$ spectrum is studied 
using the~\mbox{$\decay{\Bs}{ \jpsi \pip \pim \Pphi}$} decays. 
The~\mbox{$\decay{\Bs}{\jpsi \pip \pim \Pphi}$} candidates 
are determined with two-dimensional  unbinned extended maximum-likelihood
fit to the~$\jpsi \pip \pim \Km \Kp$ and $\Kp \Km$ mass distributions. 
The~background-subtracted  $\jpsi\Pphi$ mass spectrum 
of selected \mbox{ $\decay{\Bs}{ \jpsi \pip \pim \Pphi}$} decays 
shown in figure~\ref{fig:signal_fit6}. 
It shows a prominent structure at a mass around $4.74~\gevcc$. 
No such structure
is seen if the $\Kp \Km$ mass is restricted to the region of 
\mbox{$1.06 < m_{\Kp \Km}< 1.15\gevcc$}.
This structure cannot be explained by \mbox{$\decay{\Bs}{ \chicone(3872)\Pphi}$ }
and \mbox{$\decay{\Bs}{ \psitwos \Pphi} $} decays via a narrow intermediate
$\psitwos$ and $\chicone(3872)$ resonance since contributions from 
these decays are explicitly vetoed. 
No sizeable contributions from decays via other narrow charmonium states 
are observed in the background-subtracted $\jpsi \pip \pim$ mass spectrum.
The  $\Pphi \pip \pim$ spectrum exhibits significant deviations from 
the phase-space distribution, indicating possible presence of excited $\Pphi$ states, 
referred to as $\Pphi^{*}$ states hereafter. 
The decays \mbox{$\decay{\Bs}{\jpsi \Pphi^{*}}$} via 
intermediate $\Pphi$(1680), $\Pphi$(1850) or $\Pphi$(2170) 
states~\cite{PDG2020} 
are studied using simulated samples and no peaking structures 
are observed. 
Under the~assumption that the~observed structure, referred to as X(4740) hereafter, 
has a~resonant nature, its mass and width are determined through 
an unbinned extended maximum-likelihood fit to the~background-subtracted
$\jpsi\Pphi$ mass distribution in the~range $4.45 < m_{\jpsi\Pphi}<4.90 \gevcc$.
The fit result is superimposed in figure~\ref{fig:signal_fit6}. 
The obtained signal yield is $175\pm 39$ events 
and corresponds to a~statistical significance 
above the 5.3 $\sigma$.
The mass and width for the $\PX(4740)$~state are found to be
\begin{linenomath}
\begingroup
\allowdisplaybreaks
\begin{eqnarray*}
m_{X(4740)}&= & 4741 \pm \phantom{0}6 \pm \phantom{0}6 \gevcc\,,\\ 
\Gamma_{X(4740)}&= &\phantom{00}53 \pm 15 \pm 11 \mev.
\end{eqnarray*}
\endgroup
\end{linenomath}
 The observed parameters qualitatively agree 
 with those of the $\chicone(4700)$ state observed 
 by the LHCb collaboration in an amplitude analysis 
 of \mbox{$\decay{\Bp}{ \jpsi \Pphi \Kp}$ } 
 decays~\cite{LHCb-PAPER-2016-019,LHCb-PAPER-2016-018}.
The~obtained mass also agrees with 
the~one expected for the~$2^{++}$ \cquark\squark\cquarkbar\squarkbar tetraquark state~\cite{Ebert:2008kb}.
 
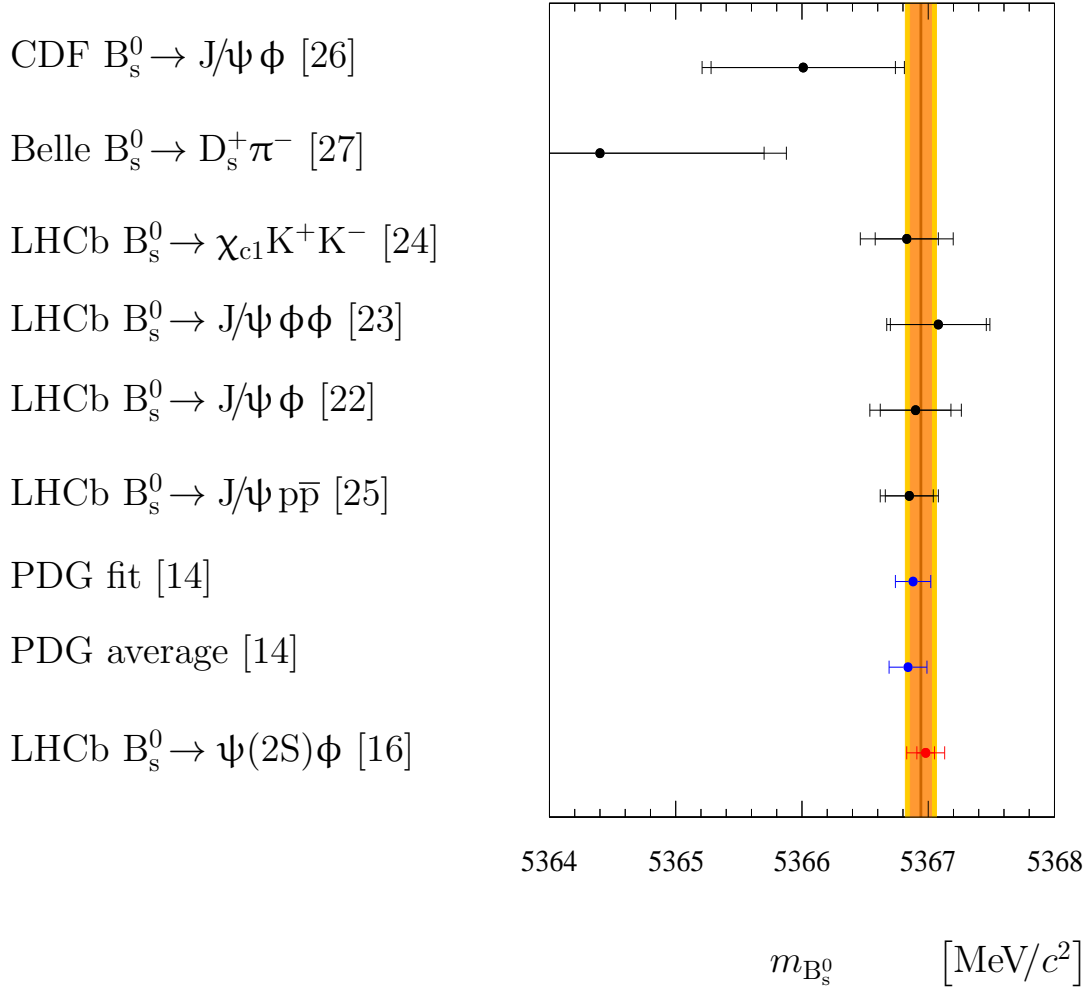
\begin{figure}[htb]
  \setlength{\unitlength}{1mm}
  \centering
  \begin{picture}(150,140)
    \put( 60, 5){ 
      \includegraphics*[height=130mm,width=80mm,%
      ]{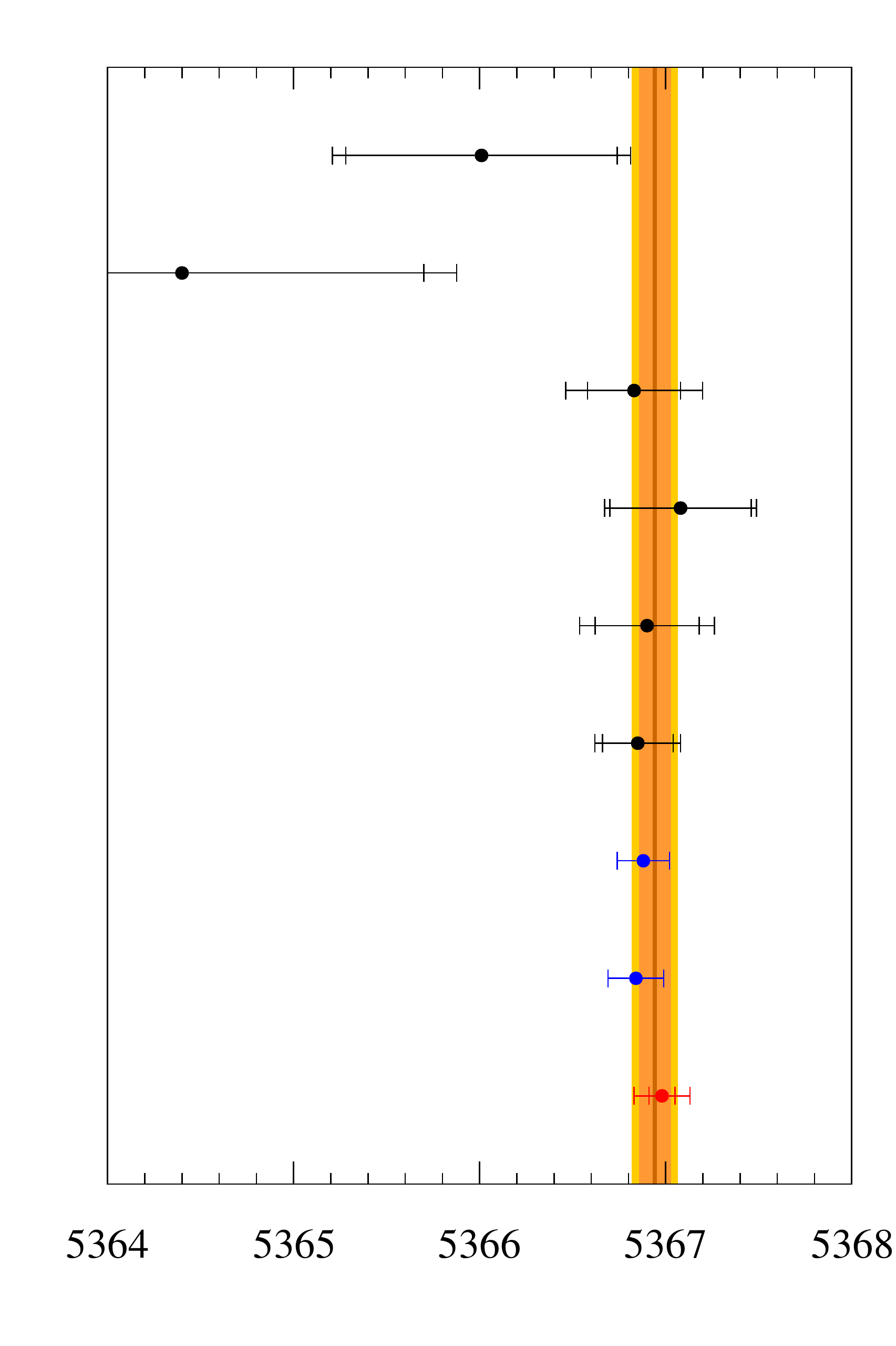}
    }  
    \put(0,120.00){\large\cdf$\decay{\Bs}{\jpsi\Pphi}$~\cite{Acosta:2005mq}}
    \put(0, 107.6){\large\belle$\decay{\Bs}{\Ds\pim}$~\cite{Louvot:2008sc}}
    \put(0, 95.40){\large\lhcb$\decay{\Bs}{\chicone\Kp\Km}$~\cite{LHCb-PAPER-2018-018}}
    \put(0, 85.0){\large\lhcb$\decay{\Bs}{\jpsi\Pphi\Pphi}$~\cite{LHcb-PAPER-2015-033}}
    \put(0, 74.40){\large\lhcb$\decay{\Bs}{\jpsi\Pphi}$~\cite{LHCb-PAPER-2011-035}}
    \put(0, 62){\large\lhcb$\decay{\Bs}{\jpsi\proton\antiproton}$~\cite{LHCb-PAPER-2018-046}}
    \put(0, 51){\large{PDG fit}~\cite{PDG2020}}
    \put(0, 41){\large{PDG average}~\cite{PDG2020}}
    \put(0, 28){\large\lhcb$\decay{\Bs}{\psitwos\Pphi}$~\cite{LHCb-PAPER-2020-035} }
    \put(100, 0){\large{$m_{\Bs}$}}
    \put(122, 0){\large{$\left[\!\mevcc\right]$}}
  \end{picture}
  \caption { \small
    Compilation of the~measurements of the~\Bs~meson mass.
     The~inner error bars indicate the~statistical uncertainty, 
     and the~outer error bars 
     correspond to quadratic sum of statistic and systematic uncertainties.
  The~band represents 
  the~value and the~uncertainty on 
  the~average  of  LHCb~measurements.
  }
  \label{fig:mass_aver}
\end{figure}

The~$\Bs$ decays to the~$\jpsi \pip \pim \Kp \Km$ final states characterize 
the~relatively small energy release allowing precise measurement of 
the~\Bs~meson mass, 
The~mass of the~$\Bs$  meson is determined from 
an~unbinned extended maximum-likelihood
fit to the $\psitwos\Kp \Km$ mass distribution for a sample of 
\mbox{$\decay{\Bs}{\jpsi \pip \pim \Kp \Km}$} decays
with $m_{\Kp \Km}<1.06 \gevcc$ and 
with the $\jpsi \pip \pim $ mass within 
a narrow region around the known mass of the $\psitwos$ meson. 
The improvement in the \Bs~mass resolution and 
significant decrease of  the systematic uncertainties 
is achieved 
by imposing a constraint on the reconstructed mass of the  $\jpsi\pip\pim$~system 
to a~known mass of the~\psitwos~meson~\cite{PDG2020,Anashin:2015rca}. 
 The measured value of the~\Bs~meson mass is found to be
 \begin{linenomath}
 \begin{equation*}
 m_{\Bs}=5366.98 \pm 0.07 \pm 0.13 \mevcc\,,     
 \end{equation*}
 \end{linenomath}
 that is the most precise single measurement of this quantity.
 This result is combined with other 
precise measurements 
by the~LHCb collaboration using~\mbox{$\decay{\Bs}{\jpsi\Pphi}$}~\cite{LHCb-PAPER-2011-035},
\mbox{$\decay{\Bs}{\jpsi\Pphi\Pphi}$}~\cite{LHcb-PAPER-2015-033}, 
\mbox{$\decay{\Bs}{\chictwo\Kp\Km}$}~\cite{LHCb-PAPER-2018-018} and 
\mbox{$\decay{\Bs}{\jpsi\proton\antiproton}$}~\cite{LHCb-PAPER-2018-046}~decays. 
The~combined mass is calculated 
accounting for correlations of systematic uncertainties between 
the~measurements. 
The~LHCb average for the~mass of the~\Bs~meson is found to be 
\begin{linenomath}
\begin{equation*} 
m_{\Bs}^{\mathrm{LHCb}} = 5366.94 \pm 0.08 \pm 0.09 \mevcc\,.
\end{equation*} 
\end{linenomath}
 The comparison with previous measurements is presented in figure~\ref{fig:mass_aver}. 

\section{Conclusions}
A~study of $\B$\nobreakdash-meson decays 
\mbox{$\decay{\Bp}{\jpsi \pip \pim \Kp}$} and 
\mbox{$\decay{\Bs}{\jpsi\pip\pim\Kp\Km}$} 
is made using $\proton\proton$~collision data 
corresponding to an~integrated luminosity of 1, 2 and 6\invfb, 
collected with the~LHCb detector at centre\nobreakdash-of\nobreakdash-mass  
energies of 7, 8 and 13\tev, 
respectively~\cite{LHCb-PAPER-2020-009,LHCb-PAPER-2020-035}.
The~reported results include 
the~first observation of the non-zero width of 
the~$\chicone(3872)$~state; 
the most precise measurement of 
the~masses of the~$\chicone(3872)$ and $\Ppsi_2(3823)$~states; 
the~first observation of 
the~\mbox{$\decay{\Ppsi_2(3823)}{\jpsi\pip\pim}$},
\mbox{$\decay{\Bu}{\Ppsi_2(3823)\Kp}$},
\mbox{$\decay{\Bs}{\chicone(3872)\left(\Kp\Km\right)_{\text{non-}\Pphi}}$}
and \mbox{$\decay{\Bs}{\jpsi\Kstarz\Kstarzb}$}~decays; 
the~most precise measurement of the~ratios of  
branching fractions of  the~\Bu and \Bs~mesons into 
the~final states with $\chicone(3823)$ and $\Ppsi_2(3823)$~particles; 
the~most precise single measurement of the \Bs~meson mass
and an~observation of a~new structure, denoted as a~$\PX(4740)$~state,
in the~$\jpsi\Pphi$~mass spectrum.


\section*{Acknowledgments}

I would like to express my gratitude to the QCD20 organizers for the great conference. Also, I'm thankful for my colleagues from LHCb collaboration who helped with preparation of this talk.






\usetikzlibrary{patterns}

\addcontentsline{toc}{section}{References}
\bibliographystyle{LHCb}
\bibliography{main,standard,LHCb-PAPER,LHCb-CONF,LHCb-DP,LHCb-TDR}
 


\end{document}